\documentclass[aps,prl,twocolumn,notitlepage,10pt,longbibliography]{revtex4-2}
\usepackage{graphicx}
\usepackage{amsfonts,amsmath,amssymb}
\usepackage[colorlinks=true]{hyperref}

\begin{document}
\title{Conformal maps and edge mode attenuation on imperfect boundaries}

\author{Grigor Adamyan}
\affiliation{
Department of Physics and Astronomy, 
Johns Hopkins University,
Baltimore, MD 21218, USA
}


\begin{abstract}




We developed a conformal map technique to analyze the attenuation of edge modes propagating along imperfect boundaries. In systems where the potential energy exhibits conformal invariance, the conformal transformation can straighten the boundary, simplifying the boundary conditions. Using the example of edge modes in a simple field-theoretical model, we examined scattering into the bulk and identified conditions that ensure the robustness of edge modes against damping. This technique has the potential to be applied to other edge-mode problems in 2+1 dimensions.

\end{abstract}

\maketitle

\newcommand{\oleg}[1]{\textcolor{red}{[#1 ---OT.]}}

\newcommand{\basti}[1]{\textcolor{blue}{[#1 ---B.]}}
\newcommand{\grigor}[1]{\textcolor{cyan}{[#1 ---G.]}}

Edge modes are fundamental in condensed matter physics, illustrating the interplay between topology, symmetry, and geometry. In the quantum Hall effect, chiral edge states enable unidirectional electron flow along sample boundaries, protected against backscattering by disorder due to their topological nature \cite{Halperin:1982}. Similarly, topological insulators host edge states with spin-momentum locking, ensuring robust transport properties protected by time-reversal symmetry \cite{Hasan:2010}. Majorana edge modes, localized at the interfaces of topological superconductors, hold promise for applications in fault-tolerant quantum computation \cite{Kitaev2001}.

However, some edge modes can be susceptible to scattering from imperfections at boundaries, which compromises their stability and propagation. This challenge has been tackled through various theoretical and numerical approaches. Numerical simulations, such as tight-binding models and finite-difference methods, provide detailed insights into the effects of roughness, defects, and disorder on edge mode propagation \cite{Groth2014}. Additionally, scattering matrix formalism further quantifies the transmission and reflection of edge modes, capturing the impact of boundary imperfections on mode stability and energy dissipation \cite{Datta1995}. These approaches highlight the importance of understanding edge mode robustness against damping.


Conformal transformations provide a complementary perspective for addressing these challenges. Widely applied in critical phenomena, conformal symmetry enables precise predictions of universal scaling behavior \cite{Cardy1996}, while in quantum field theory, it provides exact solutions for otherwise intractable systems \cite{belavin1984infinite, polyakov1970conformal, DiFrancesco1997}. In electrostatics, conformal maps are well-known for transforming complex geometries into simpler ones, facilitating the solution of Laplace's equation \cite{landau8chapter}. 
Similarly, they have been applied to zero edge modes that preserve conformal invariance \cite{Sun2012}. However, their potential for studying edge modes lacking conformal invariance remains underexplored.




In this paper, we introduce a conformal map technique to study edge modes propagating along imperfect boundaries. As an example, we analyze edge modes in a simple, physically motivated field-theoretical model with a conformally invariant potential energy. Directly solving boundary conditions for arbitrary edges poses significant challenges. To address this, we use conformal maps to transform imperfect boundaries into straight ones, simplifying the boundary conditions. This approach enables us to derive edge-mode solutions for general boundary shapes and estimate their damping. Conformal transformations for damping analysis have been previously mentioned in \cite{Dong2023}.

To begin, we consider a simple complex scalar field theory in 2+1 dimensions, which includes kinetic and potential energy densities:
\begin{equation}\label{eq:Lagrangian_density}
    \begin{split} 
        \mathcal{L} = \mathcal{K} - \mathcal{U} - \mathcal{U}_\text{top} = \frac{\rho}{2} \, \partial_t \phi \, \partial_t \bar{\phi} - \frac{\mu}{2} \, \nabla \phi \, \nabla \bar{\phi} \\
        -\frac{\Delta \mu}{2i} \left( \partial_x \bar\phi \, \partial_y \phi - \partial_x \phi \, \partial_y \bar\phi \right), 
    \end{split} 
\end{equation} where the last term is referred to as the topological term. Here, $\phi$ and $\bar{\phi}$ are complex conjugated fields, while $\rho$ and $\mu$ represent the inertia density and stiffness, respectively. The dimensionless parameter $\Delta$ characterizes the strength of the topological term, which does not contribute to the bulk equation of motion but plays a key role in the emergence of edge modes. As a physical example, recent work \cite{Pradenas:edge_modes} demonstrated that this model describes helical edge modes in a triangular antiferromagnet. In this context, $\Delta$ represents the difference in the strengths of various superexchange paths within the antiferromagnet. Alternatively, this parameter could also arise through other mechanisms.

To analyze the model (\ref{eq:Lagrangian_density}), we derive its bulk and edge modes. The bulk equation of motion is
\begin{align}
    \begin{split}
        \partial_t^2 \phi - c^2 \nabla^2 \phi = 0,
    \end{split}
\end{align}
where $c=\sqrt{\mu/\rho}$. It describes the 2-fold degenerate linear waves that propagate with velocity $c$:
\begin{align}\label{eq:spectrum_bulk_modes}
    \begin{split}
        \omega^2 = c^2 \mathbf{k}^2,
    \end{split}
\end{align}
where $\omega$ and $\mathbf{k}$ are their frequency and the wave vector, respectively.

\begin{figure}[t!]
  \centering
  \includegraphics[width=1.0\linewidth]{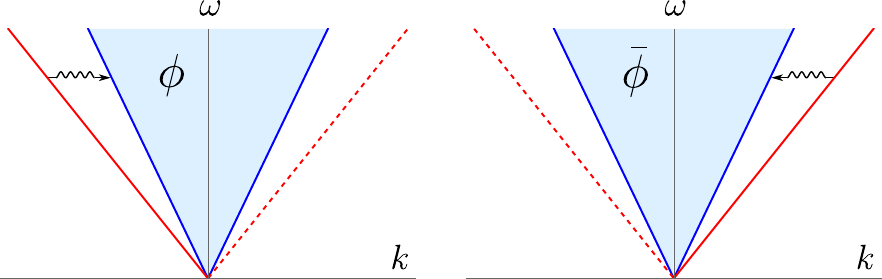}
  \caption{Spectrum of the bulk and edge modes: clockwise polarization $\phi$ (left) and counterclockwise polarization $\bar{\phi}$ (right). The blue cone corresponds to the circularly polarized waves in the bulk (\ref{eq:spectrum_bulk_modes}). The red lines represent the edge modes with dispersion (\ref{eq:spectrum_edge_mode}), where $\Delta = 0.64$. The solid red line is associated with the edge mode localized at the lower boundary, while the dashed line pertains to the upper boundary. The wavy arrows illustrate the scattering processes of the edge modes into the continuum of bulk modes.}
  \label{fig:spectrum}
\end{figure}

To identify the edge modes, we must specify a boundary condition. For the simple case of a straight boundary at $y=0$, with material occupying the upper half-plane, the boundary condition is given by:
\begin{align}\label{eq:boundary_cond}
    \begin{split}
        \partial_y \phi + i \Delta  \partial_x \phi = 0.
    \end{split}
\end{align}
By solving the bulk equation of motion with this boundary condition, we find an edge mode that is exponentially localized near the boundary:
\begin{equation}\label{eq:edge_mode_sol}
    \phi(t,x,y) = \phi_0 = \Phi \, e^{- i \omega t + i k x - \kappa y},
\end{equation}
where $\Phi$ is the amplitude of the mode, $k$ is a wave vector, and $\kappa = -\Delta k > 0$ represents the inverse localization length. The dispersion relation for the edge modes is given by:
\begin{align}\label{eq:spectrum_edge_mode}
    \omega^2 = (1-\Delta^2) c^2 k^2, 
\end{align}
indicating that they propagate at a reduced speed $c_\text{e} = \sqrt{1-\Delta^2} \, c$ compared to the bulk modes (Fig.~\ref{fig:spectrum}). 

Physically, the solution $\phi_0$ (\ref{eq:edge_mode_sol}) represents an edge mode with clockwise circular polarization propagating in a specific direction, as indicated by $\kappa > 0$. Due to the time-reversal symmetry of the system, there exists a complex conjugate solution $\bar{\phi}_0$,  which corresponds to the edge mode with anticlockwise polarization propagating in the opposite direction. These conjugated modes are therefore referred to as helical.

The edge modes derived above correspond to a material with a straight boundary. However, their behavior along imperfect boundaries raises questions about robustness against scattering. Directly solving the boundary condition for imperfect boundaries is challenging.  To overcome this, we propose using conformal maps to simplify it.

We observe that the potential part of the Lagrangian is conformally invariant, which becomes evident when the two-dimensional plane $\mathbf{r}=(x,y)$ is expressed using a complex coordinate $z = x + iy$.
Under any conformal map $z \to w(z)$, the potential  energy densities $\mathcal{U}$ and $\mathcal{U}_\text{top}$ transform as follows:
\begin{align}
    \begin{split}
        \nabla \phi \, \nabla \bar{\phi} &
        \rightarrow |\hat{J}| \, \nabla \phi \, \nabla \bar{\phi}, \\
        \left( \partial_x \bar\phi \, \partial_y \phi - \partial_x \phi \, \partial_y \bar\phi \right) &
        \rightarrow |\hat{J}| \, \left( \partial_x \bar\phi \, \partial_y \phi - \partial_x \phi \, \partial_y \bar\phi \right),
    \end{split}
\end{align}
where $|\hat{J}| = \partial_z w \, \partial_{\bar{z}} \bar{w} $ is the Jacobian of the map. Meanwhile, an infinitesimal area transforms as
\begin{align}\label{eq:conf_tranform_area}
     d^2 r \rightarrow |\hat{J}|^{-1}  d^2 r.
\end{align}
Thus, the potential energy, including the topological term, is given by:
\begin{align}
    U = \int \left(\mathcal{U} + \mathcal{U}_\text{top}\right) d^2 r,
\end{align}
and is conformally invariant. In contrast, the kinetic energy is not conformally invariant because the kinetic energy density $\mathcal{K}$ remains unchanged under a conformal map, while the infinitesimal area transforms as described in (\ref{eq:conf_tranform_area}). However, when  $\Delta$ is close to $1$,  the edge mode frequency $\omega$ approaches zero (\ref{eq:spectrum_edge_mode}), and the excitation becomes a zero-mode. In this case, the kinetic term becomes negligible, allowing us to trivially derive the edge modes for any boundary using a conformal map of the solution $\phi_0$ (\ref{eq:edge_mode_sol}). For arbitrary values of $\Delta$, the Lagrangian is not conformally invariant, suggesting that edge modes may experience damping along imperfect boundaries. The backscattering of $\phi_0$ to $\bar\phi_0$ is prohibited by spin-momentum locking, similar to the edge modes in the topological insulators \cite{Hasan:2010}. However, the system's symmetries allow for the excitation of bulk modes with the same polarization as $\phi_0$. To analyze the scattering process, conformal mapping can be used to transform an imperfect boundary into a straight one. This simplifies the boundary conditions, making them equivalent to those for a straight boundary (\ref{eq:boundary_cond}). Although the bulk equation of motion is modified, it remains solvable.

The general technique for finding edge modes on an imperfect boundary is as follows. An arbitrary boundary of a material can be represented by a conformal map of the upper half-plane $\Omega$ (with a straight boundary $y=0$):
\begin{align}\label{eq:conf_map_fourier}
    z \rightarrow w(z) = z + \sum\limits_q a_q e^{iqz},
\end{align}
where $w(z) = x' + iy'$. This map transforms the domain $\Omega$ into the domain $\Omega'$, making a deformation around the boundary. The boundary shape is determined by the Fourier amplitudes $a_q$:
\begin{align}\label{eq:boundary_shape}
    \begin{split}
        x'(x, y = 0) = x + \sum\limits_q \text{Re}\{a_q e^{iqx}\}, \\
        y'(x, y = 0) = \sum\limits_q \text{Im}\{a_q e^{iqx}\}.
    \end{split}
\end{align}
Suppose the material is described by the coordinates $\mathbf{r}' = (x', y')$ and has the deformed boundary (\ref{eq:boundary_shape}). Applying the inverse conformal transformation $w^{-1}: (x', y') \rightarrow (x, y)$ results in a straight boundary and alters the Lagrangian as follows:
\begin{align}\label{eq:JLagr}
    L = \int_{\Omega'} \mathcal{L}(\mathbf{r}') \, d^2 r' = \int_{\Omega} \left(|\hat{J}| \, \mathcal{K} - \mathcal{U} - \mathcal{U}_{\text{edge}} \right) \, d^2 r.
\end{align}
After the transformation, the prefactor $|\hat{J}|$ appears before the kinetic energy density, representing the ``cost" of straightening the boundary.

The modified bulk equation of motion of the Lagrangian $L$ is
\begin{align}\label{eq:EoM_J}
    |\hat{J}|\partial_t^2 \phi - c^2 \nabla^2 \phi = 0.
\end{align}
For small boundary deformations, the Jacobian slightly deviates from unity and can be approximated as
\begin{align}\label{eq:Jacobian_approx}
    |\hat{J}| = \partial_z w \, \partial_{\bar{z}} \bar{w} \approx 1 - 2 \sum\limits_{q} q \, \text{Im}\{a_{q} e^{iqz}\}.
\end{align}
This modifies the equation of motion (\ref{eq:EoM_J}) to
\begin{align}\label{eq:EoM_J_perturb}
    \partial_t^2 \phi - c^2 \nabla^2 \phi = 2 \, \partial_t^2 \phi \sum\limits_{q} q \, \text{Im}\{a_{q} e^{iqx-qy}\}.
\end{align}
Given the small magnitude of the right-hand side, perturbation theory is applied to obtain the solution in the form $\phi = \phi_0 + \delta \phi$.  This approach leads to a linear wave equation for $\delta \phi$ with a source term:
\begin{align}\label{eq:EoM_perturb_source_term}
    \partial_t^2 \delta\phi - c^2 \nabla^2 \delta\phi = i \omega^2 \phi_0 \sum\limits_{\pm, q} \pm q \, a^{\pm}_{q} e^{\pm iqx - qy},
\end{align}
where the notation $a_{q}^{+} = a_{q}$ and $a_{q}^{-} = \bar{a}_{q}$ is introduced for convenience. The solution of the inhomogeneous equation is given by:
\begin{align}
    \delta \phi_{\text{A}} = \sum\limits_{\pm, q} A_{q}^{\pm} e^{-i \omega t + i(k \pm q)x - (q-\Delta k)y},
\end{align}
with coefficients
\begin{align}
    A_{q}^{\pm} = \frac{i}{2} \Phi (1 \mp \Delta) k a_{q}^{\pm}.
\end{align}
This represents additional spin waves localized near the boundary. To satisfy the boundary condition (\ref{eq:boundary_cond}), we must include the solution of the homogeneous equation:
\begin{align}\label{eq:delta_phi_B}
    \delta \phi_{\text{B}} = \sum\limits_{\pm, q} B_{q}^{\pm} e^{-i \omega t + i(k \pm q)x - \kappa_{q}^{\pm} y},
\end{align}
where the coefficients $B_{q}^{\pm}$ and $\kappa_{q}^{\pm}$ are derived by using the bulk equation of motion (\ref{eq:EoM_perturb_source_term}):
\begin{align}
    \begin{split}
        B_{q}^{\pm} = -A_{q}^{\pm} \frac{q(1 \pm \Delta)}{\Delta(k \pm q) + \kappa_{q}^{\pm}}, \\
        \left( \kappa_{q}^{\pm} \right)^2 = q^2 \pm 2kq + \Delta^2 k^2.
    \end{split}
\end{align}
If $\left( \kappa_{q}^{\pm} \right)^2 < 0$, such terms in the solution $\delta \phi_{\text{B}}$ correspond to radiated waves from the boundary. This condition is satisfied when
\begin{align}\label{eq:emission_kappa_y}
    |q \pm k| \leq |k| \sqrt{1-\Delta^2}.
\end{align}
\begin{figure}[t!]
  \centering
  \hspace{-2.5em}
  \includegraphics[width=0.95\linewidth]{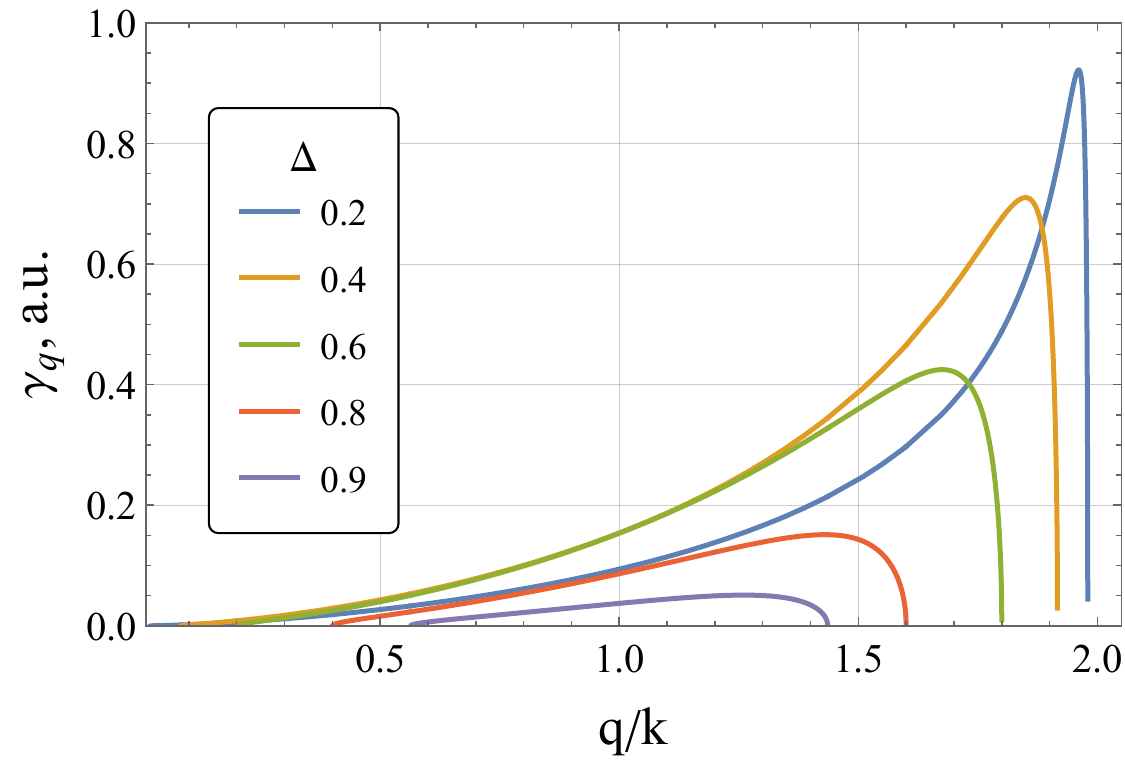}
  \caption{The density of the damping factor $\gamma_q$ from $q$ varies with different values of $\Delta$. It illustrates how damping disappears as $\Delta$ approaches one, indicating that the system approaches conformal invariance.}
  \label{fig:damping}
\end{figure}
It indicates that a non-straight boundary leads to elastic scattering of the edge mode, exciting the bulk modes, as shown in Fig.~\ref{fig:spectrum}. Conversely, when $\left( \kappa_{q}^{\pm} \right)^2 > 0$, the terms in $\delta \phi_{\text{B}}$ correspond to waves that remain localized near the boundary. Together with the solution $\delta \phi_{\text{A}}$, these waves only slightly adjust the localization length $\kappa^{-1}$ of the edge mode.

Finally, by combining the solutions of the inhomogeneous and homogeneous equations, the overall solution can be expressed as
\begin{align}
    \phi = \phi_0 + \delta \phi_{\text{A}} + \delta \phi_{\text{B}}.
\end{align}

Using the developed technique, we can analyze the robustness of the edge mode against scattering and estimate the damping for a general boundary shape defined by the conformal map (\ref{eq:conf_map_fourier}). The solution of the homogeneous equation, $\delta \phi_\text{B}$, includes a set of emitting waves that carry away the energy of the edge mode to the bulk. The attenuation can be calculated by averaging the energy flow in the $y$-direction over the boundary, which yields
\begin{align}
    \begin{split}
        \langle j_y \rangle = - \mu \, \langle \text{Re} \{ \partial_t \phi \, \partial_y \bar\phi \} \rangle = \mu \omega \sum\limits_{\pm, q} |B_{q}^{\pm}|^2  \, \text{Im} \{ \kappa_{q}^{\pm} \}.
    \end{split}
\end{align}
Each term in the summation that satisfies Eq. (\ref{eq:emission_kappa_y}) contributes additively to the damping factor, defined as the relative energy loss averaged over one period:
\begin{align}
    \gamma = \sum\limits_{q} \gamma_q |a_q|^2,
\end{align}
where $\gamma_q$ is the density of the damping factor, which depends on $q$. The dependence of $\gamma_q$ on different values of $\Delta$ is shown in Fig.~\ref{fig:damping}. An upper bound for the total damping can be expressed as
\cite{supmat}:
\begin{align}
    \gamma \leq (ka)^2 (1-\Delta^2)^2,
\end{align}
where $a$ is the root mean square depth of the boundary imperfection. This result highlights the dependence of damping on the shape of the boundary deformations as well as the parameter $\Delta$. 

As $\Delta$ approaches 1, damping is suppressed, indicating the system's convergence to conformal invariance, as shown in Fig.~\ref{fig:damping}. Additionally, damping becomes negligible for long wavelengths of the edge mode ($ka \ll 1$). This can be explained as follows: at long wavelengths, the penetration length into the bulk, $\kappa^{-1}$, becomes large, allowing the energy of the edge mode to spread far from the boundary.  As a result, the relative energy loss near the boundary becomes insignificant.




In addition to boundary imperfections, any material may have corners where the boundary changes direction. Let $\sigma$  represent the characteristic length scale of a corner. Then, the function $a_q$ (\ref{eq:conf_map_fourier}) has a characteristic width $\sigma^{-1}$. 
The emitted waves from the boundary correspond to Fourier harmonics $a_q$ with $q$ comparable to the edge mode wave vector $k$ (\ref{eq:emission_kappa_y}). For short wavelengths where $k \gg \sigma^{-1}$, such harmonics $a_{q}$ become negligible, effectively suppressing the damping of the edge mode. The extent of suppression depends on how rapidly $a_q$ decays at large $q$, which is dictated by the shape of the boundary.

\begin{figure}[t!]
  \centering
  \includegraphics[width=0.98\linewidth]{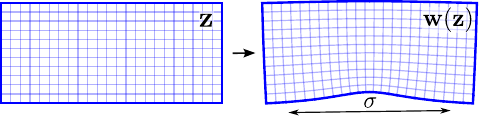}
  \caption{An example of a conformal map $w(z)$ (\ref{eq:conf_map_curved_bound}) that transforms a straight boundary into a curved one with a bulge.}
  \label{fig:conformal_map_w4}
\end{figure}

To analyze this quantitatively, we modeled the corner as a bulge on the boundary, avoiding issues related to the Jacobian divergences in the complex plane. Let us consider an example of a bulge defined by the conformal map (Fig.~\ref{fig:conformal_map_w4}):
\begin{align}\label{eq:conf_map_curved_bound}
    w(z) = z - \frac{b}{z/\sigma + i},
\end{align}
where $b$ and $\sigma$ represent the characteristic depth and width of the bulge, respectively. The resulting boundary shape from this conformal map is expressed as
\begin{align}
    y'(x) = \frac{b}{1+ x^2/\sigma^2}.
\end{align}
The Fourier amplitudes $a_q$, corresponding to the decomposition in (\ref{eq:conf_map_fourier}),  are given by:
\begin{align}
    a_q = i b \sigma \, \theta(q) \, e^{-q\sigma},
\end{align}
where $\theta(q)$ is the Heaviside step function. This amplitude decays exponentially with $q$,  resulting in an exponential suppression of the damping factor: 
\begin{align}
    \gamma \propto e^{-k\sigma}.
\end{align}
Thus, it significantly protects the edge modes with short wavelengths ($k\sigma \gg 1$) from damping at corners.





 
To conclude, we studied the edge modes in the simple model (\ref{eq:Lagrangian_density}) that propagate along imperfect boundaries. The central result of this work is the development of a conformal mapping technique that transforms an imperfect boundary into a straight one, enabling the analysis of edge modes on arbitrary boundaries. Our findings reveal that the edge modes hybridize with bulk modes, leading to a finite lifetime. By analyzing the damping factor, we determined that the robustness of the edge modes against scattering depends on their wavelengths, $\lambda$. These wavelengths must lie within the range defined by the root mean square depth of boundary imperfections, $a$, and the characteristic length scale of a corner, $\sigma$:
\begin{align}
    a \ll \lambda \ll \sigma.
\end{align}
Experimentally, this suggests a potentially broad range of $\lambda$ values, extending from the interatomic scale to the size of the material.



The conformal mapping technique can also be applied to other edge mode problems in 2+1 dimensions, provided the potential energy preserves conformal invariance. Notable examples satisfying this condition include Rayleigh waves \cite{LL7} and Dyakonov electromagnetic surface waves in composite materials \cite{D’yakonov1988}.






I am deeply grateful to Oleg Tchernyshyov for his guidance and support throughout this project. I thank Ibrahima Bah, Leonid Levitov, Bastian Pradenas, and Gurgen Adamian for helpful discussions. I sincerely appreciate Mariia Kryvoruchko for reading the paper and for her valuable comments. The author would like to acknowledge the William H. Miller III Graduate Fellowship for supporting this research.

\vspace{1em}

\bibliography{references}



\begin{widetext}
\section*{Supplementary Material}
We investigate the upper bound of edge mode damping in materials with arbitrarily imperfect boundaries. The expression for the average energy flow is given by
\begin{align}
    \langle j_y \rangle = - \mu \, \langle \text{Re} \{ \partial_t \phi \, \partial_y \bar\phi \} \rangle = \mu \omega \sum\limits_{\pm, q} |B_{q}^{\pm}|^2 \, \text{Im} \{ \kappa_{q}^{\pm} \}.
\end{align}
For such terms corresponding to the emitted waves (\ref{eq:emission_kappa_y}), we have:
\begin{align}
    |B_{q}^{\pm}|^2 \text{Im} \{ \kappa_{q}^{\pm} \} = \mp \frac{1}{4}|\Phi|^2 k^2 (1-\Delta^2) q |a_q|^2 \, \xi_{\Delta}\left( \pm \frac{q}{k} \right),
\end{align}
where the function $\xi_\Delta(x)$ is defined as
\begin{align}
    \xi_\Delta(x) = \frac{\sqrt{-(x^2 + 2x + \Delta^2)}}{2+x}.
\end{align}
This function attains its maximum at the point $x_0$, where $\partial_x \xi_\Delta(x_0) = 0$:
\begin{align}
    \begin{split}
         x_0 &= \Delta^2 - 2, \\
         \xi_\Delta(x_0) &= \frac{1-\Delta^2}{|\Delta|}.
    \end{split}
\end{align}
Additionally, using the condition $|q|<2|k|$ derived from (\ref{eq:emission_kappa_y}), we establish an upper bound for the energy flow:
\begin{align}\label{eq:j_upper_bound}
    \langle j_y \rangle \leq \mu c \, |\Phi|^2 k^4 a^2 \frac{(1-\Delta^2)^{5/2}}{|\Delta|},
\end{align}
where
\begin{align}
    a^2 = \sum\limits_{q} |a_q|^2.
\end{align}
According to Parseval's identity, $a^2$ is the mean-squared depth of the boundary roughness $\langle (y')^2 \rangle$.

The energy density corresponds to the zero component of Noether's current, which is associated with time-translation symmetry. For the edge mode, the average energy density is expressed as
\begin{align}
    \langle \mathcal{E} \rangle = \mu |\Phi|^2 \frac{|k|}{2|\Delta|}.
\end{align}
The damping factor, which quantifies the relative energy loss per period, is given by
\begin{align}
    \gamma = \frac{\langle j_y \rangle}{2 \omega \langle \mathcal{E} \rangle}.
\end{align}
Using the estimated energy flow $\langle j_y \rangle$ from (\ref{eq:j_upper_bound}), the upper bound for the damping factor is
\begin{align}
    \gamma \leq (ka)^2 (1-\Delta^2)^2.
\end{align}

\end{widetext}

\end{document}